\documentclass[american,spanish,english,aps,preprint]{revtex4}
\usepackage[T1]{fontenc}
\usepackage[latin9]{inputenc}
\usepackage{amstext}
\usepackage{graphicx}
\usepackage{amssymb}

\makeatletter

\providecommand{\LyX}{L\kern-.1667em\lower.25em\hbox{Y}\kern-.125emX\@}

\@ifundefined{textcolor}{}
{%
 \definecolor{BLACK}{gray}{0}
 \definecolor{WHITE}{gray}{1}
 \definecolor{RED}{rgb}{1,0,0}
 \definecolor{GREEN}{rgb}{0,1,0}
 \definecolor{BLUE}{rgb}{0,0,1}
 \definecolor{CYAN}{cmyk}{1,0,0,0}
 \definecolor{MAGENTA}{cmyk}{0,1,0,0}
 \definecolor{YELLOW}{cmyk}{0,0,1,0}
 }

\makeatother

\usepackage{babel}
\addto\shorthandsspanish{\spanishdeactivate{~<>}}

\begin{document}
\selectlanguage{spanish}%

\preprint{\global\long\def\abs#1{\left| #1 \right| }
\global\long\def\ket#1{\left| #1 \right\rangle }
\global\long\def\bra#1{\left\langle #1 \right| }
\global\long\def\half{\frac{1}{2}}
\global\long\def\partder#1#2{\frac{\partial#1}{\partial#2}}
\global\long\def\comm#1#2{\left[ #1 ,#2 \right] }
\global\long\def\vp{\vec{p}}
\global\long\def\vpp{\vec{p}\, ^{\prime}}
\global\long\def\dt#1{\delta^{(3)}(#1 )}
\global\long\def\Tr#1{\textrm{Tr}\left\{  #1 \right\}  }
\global\long\def\Real#1{\mathrm{Re}\left\{  #1 \right\}  }
\global\long\def\braket#1{\langle#1\rangle}
}

\selectlanguage{english}%

\preprint{IFIC/09-57 and FTUV-09-1111}

\title{Information of a qubit interacting with a multilevel environment}

\author{A. Pérez}

\affiliation{Departament de Física Teòrica and IFIC, Universitat de València-CSIC
\\
 Dr. Moliner 50, 46100-Burjassot, Spain}
\begin{abstract}
We consider the interaction of a small quantum system (a qubit) with
a structured environment consisting on many levels. The qubit will
experience a decoherence process, which implies that its initial information
will be transferred to the environment. We investigate how this information
is distributed on a given subset of levels as a function of its size,
using the mutual information between both entities, in the spirit
of the partial information plots studied by Zurek \cite{Zurek2009}.
In this case we observe some differences, which arise from the fact
that we are partitioning just one quantum system, and not a collection
of them. However some similar features, like a redundancy (in the
sense that a given amount of information is shared by many subsets),
which increases with the size of the environment, are also found here. 
\end{abstract}
\maketitle

\section{introduction}

Quantum systems are usually subject to interaction with some environment.
Such interaction is at the origin of the decoherence experienced by
the system \cite{Breuer2007,Weiss2008}, a fact that makes difficult
the design and performance of quantum computers \cite{nielsen:2000}.
In short, decoherence will cause a system \textit{S} to evolve from
a pure state to a mixed one, thus destroying the subtle correlations
(entanglement) needed for quantum computation and for many kinds of
experiments. During this process, the entropy of the system increases
and, as a consequence, the information it contained is transferred
to the environment \textit{E}, where it is inevitably lost.

This is, at least, the usual point of view. Suppose, however, that
one can have access to this information. If we consider the combined
\textit{S+E} total system as isolated, then the dynamics of this enlarged
system is unitary, which means that the total entropy is conserved.
In other words, all the information lost by \textit{S} must be \textbf{necessarily}
encoded on \textit{E} . One can then address the problem of knowing
how this information is stored in the environment. There are reasons
to justify this research. First, a knowledge of how the interaction
of the system with the environment works can be used to design a strategy
to protect the system against decoherence. Indeed, procedures to control
the effects caused by decoherence have been proposed, and a better
understanding of the decoherence mechanism may be used to improve
these strategies \cite{Montina2008,Uhrig07,Vink2009,Du09,Amin2008,Gordon2008}.

A second reason is related to our perception of quantum systems. As
pointed out by W. Zurek \cite{Zurek2009}, observers usually perceive
those systems by an indirect way, i.e. by interaction with the environment
to which the system is coupled. Such interaction leads to a proliferation
(redundancy) of the information content of \textit{S} within \textit{E},
a fact that can help us to understand why many independent observers
can agree about the properties of the system \textit{S} or, in other
words, how these properties can become \textit{objective}. As shown
in these references, the above mentioned redundancy comes at the price
that only some selected states can give rise to a large offspring.
This fact has been referred to as \textit{quantum Darwinism} \cite{Zurek2009,Blume-Kohout2006,Blume-KohoutPhys.Rev.Lett.101:2404052008,Ollivier2005}. 

In order to investigate these features, Zurek and his coworkers have
considered a quantum system (a qubit, for example) coupled to an environment
which consists on many identical quantum entities, such as spins or
oscillators. One can then choose an arbitrary subset of \textit{E},
and study the mutual information this particular partition shares
with \textit{S . }From these \textit{Partial Information Plots} (PIPs)
one can get insight about how many subsets of the whole environment
share a given amount of information with the smaller system \textit{S. }

In this paper, we will face a different topic, although it is related
to, and motivated by, the above discussion. We calculate the mutual
information between a qubit and a multilevel environment (a qudit).
We analyze this quantity for a fraction of the total number of levels
as a function of the size of the chosen fraction. The purpose of this
study is to investigate how many levels of the environment one should
{}``read'' in order to obtain information about the qubit, assuming
this information is available experimentally. The obtained PIPs differ
from the ones mentioned above, because in this case we are discussing
the interaction with a subset of a \textbf{single} quantum system,
instead of a collection of them. Some similarities, however, still
exist. We discuss these topics. As we also show, the fraction of levels
one should measure in order to gain a substantial information about
the system \textit{S} depends on the total number of levels of the
environment.

This paper is organized as follows. In Sect. II we introduce the model
used for the system and environment. In Sect. III we present the results
that are derived from our calculations. Sect. IV will summarize these
results. We work in units such that $\hbar=1$.

\section{model}

We use a simple model in order to describe the decoherence of our
system (the qubit). It is assumed to interact with an environment
consisting on a band of $N$ equally spaced levels. This model can
describe relaxation to equilibrium and decoherence effects in a natural
way \cite{Perez2009}, and may be regarded as a simplified version
of the two-band model described in \cite{Breuer2006,Michel2006} .
We write the Hamiltonian for the free qubit as \begin{equation}
H_{S}=\frac{\Delta E}{2}\sigma_{z},\end{equation}
where $\Delta E$ is the energy gap for our two-level system and $\sigma_{z}$
is the third Pauli matrix. The Hamiltonian describing the environment
is defined by\foreignlanguage{american}{ \begin{eqnarray}
H_{E} & = & \sum_{n=1}^{N}\frac{\delta\varepsilon}{N}n|n\rangle\langle n|\end{eqnarray}
 and the interaction between both systems by \begin{equation}
H_{I}=\sigma_{z}C,\label{MODEL-V}\end{equation}
}

\selectlanguage{american}%
with \begin{equation}
C=\lambda\sum_{n_{2}>n_{1}}c(n_{1},n_{2})|n_{1}\rangle\langle n_{2}|+h.c.\label{couplings}\end{equation}

acting on \textit{E} and $\sigma_{z}$ acting on \textit{S}. The indices
$n,n_{1}$ and $n_{2}$ label the levels of the energy band. The global
strength of the interaction with the qubit is given by $\lambda$.
The coupling constants $c(n_{1},n_{2})$ are independent Gaussian
random variables. Their averages (denoted by $<>$) over the random
constants $c(n_{1},n_{2})$ satisfy: \begin{eqnarray}
\langle c(n_{1},n_{2})\rangle & = & 0,\label{AV1}\\
\langle c(n_{1},n_{2})c(n'_{1},n'_{2})\rangle & = & 0,\label{AV2}\\
\langle c(n_{1},n_{2})c^{*}(n'_{1},n'_{2})\rangle & = & \delta_{n_{1},n'_{1}}\delta_{n_{2},n'_{2}}.\label{AV3}\end{eqnarray}

\selectlanguage{english}%

\section{Results}

We now present some results obtained from a simulation of the model
introduced in the previous section. The combined \textit{S+E} total
system is considered as isolated, starting from a pure factorizable
state $\ket{\Psi(0)}=\ket{\Psi_{S}(0)}\otimes\ket{\Psi_{E}(0)}$ ,
and we let the whole system evolve according to the Schrödinger equation,
thus obtaining $\ket{\Psi(t)}$ as a function of time from \begin{equation}
i\frac{d}{dt}\ket{\Psi(t)}=H\ket{\Psi(t)},\label{Scho}\end{equation}

with $H=H_{S}+H_{E}+H_{I}$ the total Hamiltonian. As for the initial
conditions, we choose $\ket{\Psi_{S}(0)}=\frac{1}{\sqrt{2}}\ket ++\frac{1}{\sqrt{2}}\ket -$
, where $\ket{\pm}$ are the eigenstates of $H_{S}$. For the environment,
we take the uniform superposition $\ket{\Psi_{E}(0)}=\frac{1}{\sqrt{N}}\sum_{n=1}^{N}\ket n$.
As the total system evolves in time, the qubit becomes entangled with
the environment, so that it can not be described as a pure state.
We obtain the reduced density matrix of \textit{S} as \begin{equation}
\rho_{S}(t)\equiv Tr_{E}\{\rho(t)\},\end{equation}
\begin{figure}
\includegraphics[width=8cm]{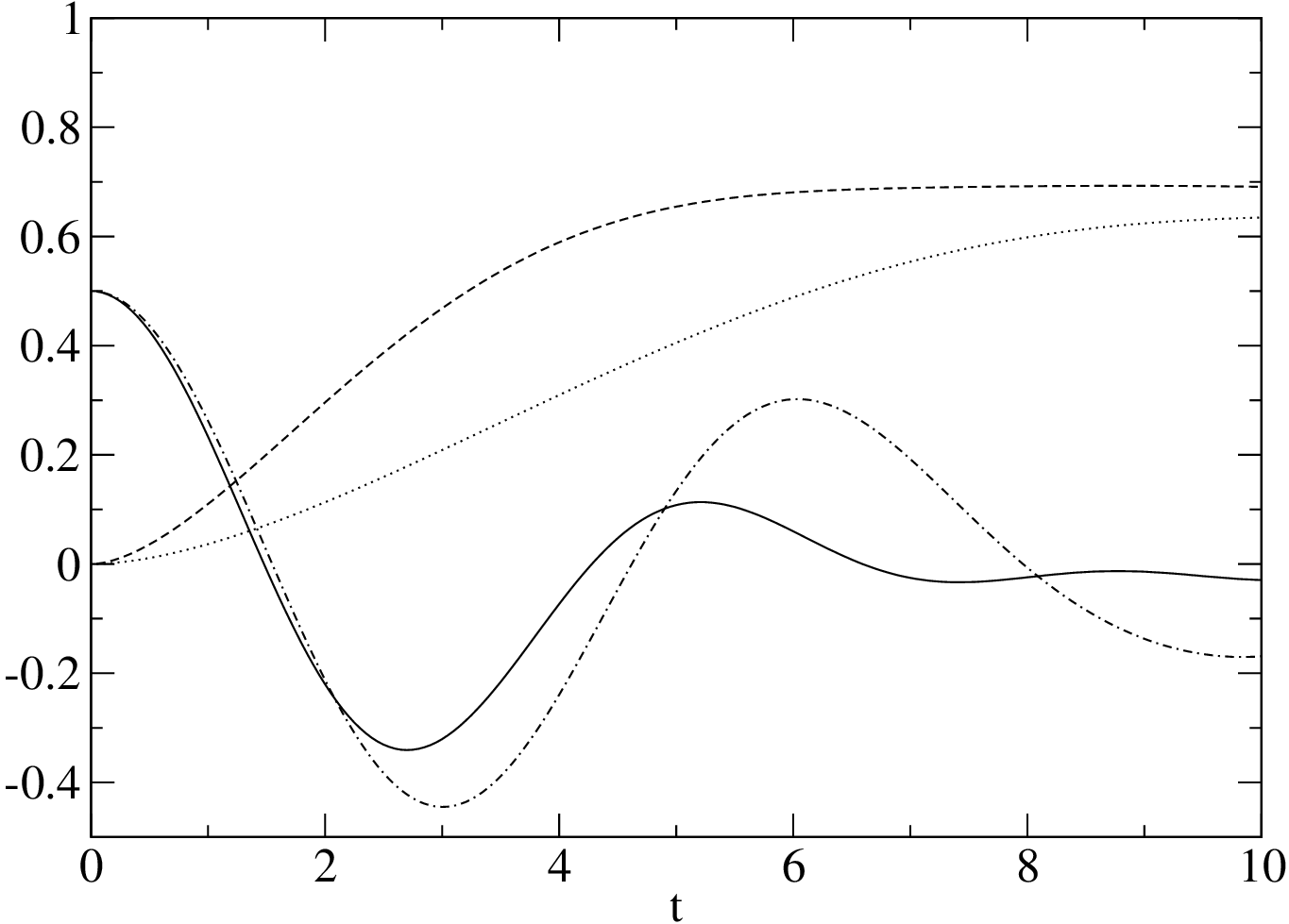}

\caption{\label{n10and100dat}Numerical solution of Eq. (\ref{Scho}) for an
environment of $N=10$ or $N=100$ levels. The dashed-dotted and continuous
lines show the evolution of $\rho_{S}(1,2)$ for $N=10$ and $N=100$
, respectively. The dotted and dashed lines are plots of the entropy
of the qubit for $N=10$ and $N=100$ , respectively.}

\end{figure}

where $\rho(t)=\ket{\Psi(t)}\bra{\Psi(t)}$ is the density matrix
corresponding to \textit{S+E} , and $Tr_{E}$ stands for the partial
trace over the environment. From this, we calculate the entropy of
system \textit{S} as $S_{S}(t)=\Tr{\rho_{S}(t)\log\rho_{S}(t)}$. 

Fig. \ref{n10and100dat} plots the results obtained from a numerical
simulation of Eq. (\ref{Scho}) when our system interacts with a small
environment with $N=10$ levels. The rest of parameters are $\lambda=2.5\times10^{-2},$
$\Delta E=1$ , $\delta\varepsilon=0.5$ . Although we solve Eq. (\ref{Scho})
exactly, this choice warranties that the evolution of $\rho_{S}(t)$
can be approximated by a master equation of the form \cite{Perez2009}

\begin{equation}
\frac{d}{dt}\rho_{S}(t)=-i[H_{S},\rho_{S}(t)]+\Gamma(\sigma_{z}\rho_{S}(t)\sigma_{z}-\rho_{S}(t)).\label{masterGrov}\end{equation}
with 

\begin{equation}
\Gamma=\frac{2\pi\lambda^{2}N}{\delta\varepsilon}.\end{equation}
It has been showed that Eq. (\ref{masterGrov}) will approximate the
evolution of $\rho_{S}(t)$ when the conditions \begin{eqnarray}
c_{1}\equiv\frac{\lambda N}{\delta\varepsilon} & \geq & \frac{1}{2}\nonumber \\
c_{2}\equiv\frac{\lambda^{2}N}{\delta\varepsilon^{2}} & \ll & 1\label{criteria}\end{eqnarray}

are met \cite{Michel2006,Gemmer2004}. In our case, we find $c_{1}=0.5,c_{2}=2.5\times10^{-2}$. 

We clearly see that the system oscillates, but these oscillations
are damped due to the effect of decoherence, which translates into
an increase of the entropy. These features can be easily reproduced
by studying Eq. (\ref{masterGrov}), and imply that the information
initially stored in the system has degraded. Since the \textit{S+E}
total system is assumed to be isolated, such information \textbf{must
be present} in the environment \textit{E.} The question we want now
to analyze is how this information is distributed or, in other words,
how much of the environment one should scan in order to known this
information. 

Let us write the time-dependent global state as\begin{equation}
\ket{\Psi(t)}=\sum_{i=1}^{2}\sum_{n=1}^{N}a_{in}(t)\ket i\otimes\ket n.\end{equation}

In this equation, $\{\ket i/i=1,2\}$ is a basis of the Hilbert space
associated to the qubit, and $a_{in}(t)$ are the coefficients of
the expansion in the composite base $\{\ket i\otimes\ket n\}$. One
then has \begin{equation}
\rho(t)=\sum_{i,j=1}^{2}\sum_{n,m=1}^{N}a_{in}(t)a_{jm}^{*}(t)\ket i\bra j\otimes\ket n\bra m.\end{equation}
Assume one can access a given subset \textit{F} of $n_{F}$ levels
(not necessarily consecutive) out of the $N$ total number of levels
in \textit{E} . The density matrix $\rho_{SF}$ corresponding to \textit{F+S}
can be obtained from \begin{equation}
\rho_{SF}(t)=\frac{1}{N_{F}}\sum_{i,j=1}^{2}\sum_{n,m\in F}a_{in}(t)a_{jm}^{*}(t)\ket i\bra j\otimes\ket n\bra m.\end{equation}

where $N_{F}=\sum_{i=1}^{2}\sum_{n\in F}|a_{in}|^{2}$ is a normalization
factor, so that $\Tr{\rho_{SF}(t)}=1$ (we have omitted the dependence
on $t$ for brevity). It can be easily checked that this density matrix
actually describes a pure state, since \begin{equation}
\rho_{SF}(t)=\ket{\Psi_{SF}(t)}\bra{\Psi_{SF}(t)},\end{equation}
with \begin{equation}
\ket{\Psi_{SF}(t)}=\sum_{i=1}^{2}\sum_{n\in F}a_{in}(t)\ket i\otimes\ket n.\end{equation}
Finally, we obtain the density matrix for \textit{F} from \begin{equation}
\rho_{F}(t)\equiv Tr_{S}\{\rho_{SF}(t)\}=\frac{1}{N_{F}}\sum_{i=1}^{2}\sum_{n,m\in F}a_{in}(t)a_{im}^{*}(t)\ket n\bra m.\end{equation}
In order to characterise how much information we can obtain from \textit{S}
by knowing about \textit{F} , we define the mutual information between
\textit{F} and \textit{S} \begin{equation}
I(S:F)=S_{S}+S_{F}-S_{SF}=S_{S}+S_{F},\label{mutinf}\end{equation}
where $S_{F}=\Tr{\rho_{F}(t)\log\rho_{F}(t)}$ is the entropy associated
to \textit{F} , and $S_{SF}=\Tr{\rho_{SF}(t)\log\rho_{SF}(t)}$ is
the entropy associated to \textit{F+S} , which vanishes according
to the above discussion. The last equality in Eq. (\ref{mutinf})
immediately follows from this. We now analyze the magnitude $I(S:F)$
as a function of the fraction $f=n_{F}/N$ of levels involved in \textit{F}.
\begin{figure}
\includegraphics[width=8cm]{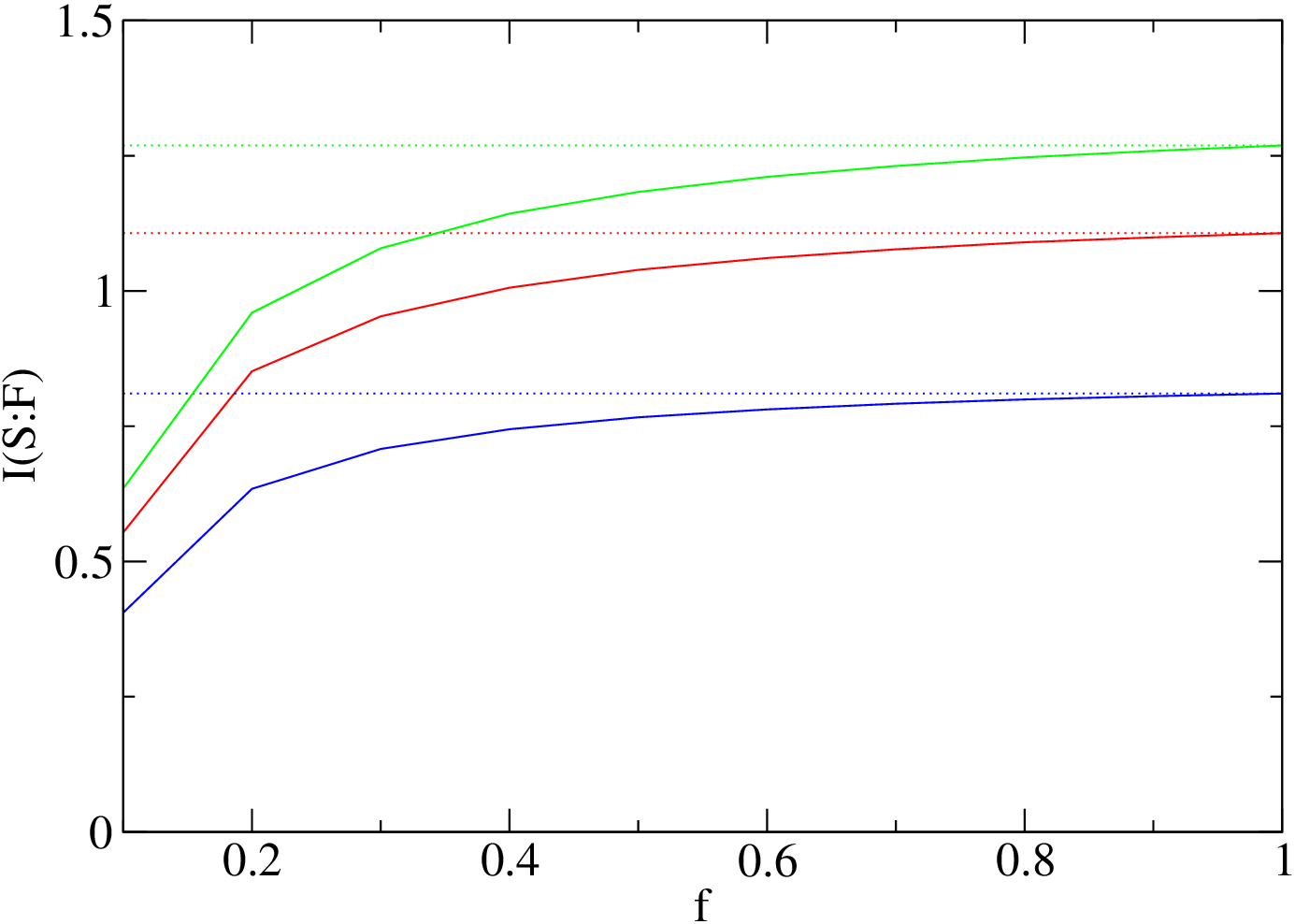}

\caption{\label{n10mutual}(Color online). Averaged mutual information (solid
lines) as a function of the fraction $f$ , corresponding to the model
with $N=10$ levels, for 3 different times: $t=5,7,10$ (from bottom
to top). The values of the model are the same as in Fig. (\ref{n10and100dat}).
For each $t$ , the horizontal dotted line shows the maximum value
$2S_{S}(t)$.}

\end{figure}

Given a value of time, we have obtained, from the numerical simulations,
that $I(S:F)$ is not a monotonic function of $f$ , i.e. the information
has accumulated in some levels at the expense of the rest. However,
if we perform an averaging over all levels participating for a given
fraction $f$ , one expects a monotonic increase. This is indeed the
case, as we show in Fig. \ref{n10mutual}. At this point, we observe
a similarity with the partial information plots studied in \cite{Zurek2009,Blume-KohoutPhys.Rev.Lett.101:2404052008,Blume-Kohout2006,Blume-Kohout2005}.
There is, however, a fundamental difference between both kind of plots,
which has to be stressed. In the previous case, the authors consider
the interaction of a qubit with an environment composed by \textbf{several}
quantum systems (like qubits or oscillators). In our case, the environment
is just one quantum system, although it consists on many levels. Partitioning
these levels is not the same as partitioning several quantum systems
into a subset of them. In other words, let $\mathbb{\mathcal{H_{N}}}$
be the Hilbert space associated to \textit{E} , $\mathcal{H_{F}}$
the Hilbert space corresponding to \textit{F} and $\mathcal{H_{\bar{F}}}$
the one associated to its complementary in \textit{E}. Obviously,
$\mathcal{H_{N}}$ is obtained as the direct sum $\mathcal{H_{N}}=\mathcal{H_{F}}\oplus\mathcal{H_{\bar{F}}}$
and \textbf{not} as the tensor product $\mathcal{H_{F}}\otimes\mathcal{H_{\bar{F}}}$,
as it would appear when the environment is made from several quantum
objects, and \textit{F} represents a subset of them. This has the
consequence, for example, that strong subadditivity \cite{nielsen:2000}
does not apply to F and its complementary. Another consequence is
that the plot of $I(S:F)$ versus $f$ is not symmetric, differently
to what is obtained by Zurek and coworkers. 

Figure \ref{n10mutual} is a plot of $I(S:F)$ as a function of $f$
for the same model considered in Fig. \ref{n10and100dat} and three
different times : $t=5,7,10$ . For each value of $n_{F}$, we have
performed an average over all possible $(\begin{array}{c}
N\\
n_{F}\end{array})$ combinations. As can be seen, the resulting curves are monotonic
functions. The maximum value is attained when $f=1$ or, equivalently,
when $n_{F}=N$ , which amounts to knowing the total information in
the environment. In this case, the partition of \textit{F+S} corresponds
to two entangled quantum systems ( \textit{E} and \textit{S} ) sharing
the same information $S_{S}(t)$. Therefore, this maximum value is
\begin{equation}
I(S:E)=2S_{S}(t).\end{equation}

We have plotted this maximum value for each time $t$ as a horizontal
dotted line, corresponding to twice the entropy of the qubit shown
in Fig. \ref{n10and100dat}. 

In order to explore a more complex environment, we have also performed
a simulation when the number of levels is $N=100$ . The results for
the time evolution of $\rho_{S}(1,2)$ and the entropy $S_{S}(t)$
are also shown in Fig. \ref{n10and100dat}. In this case we have taken
$\lambda=1.5\times10^{-2},$ the rest of parameters being the same
as in the previous example, giving $c_{1}=3,c_{2}=9\times10^{-2}$.
\begin{figure}
\includegraphics[width=8cm]{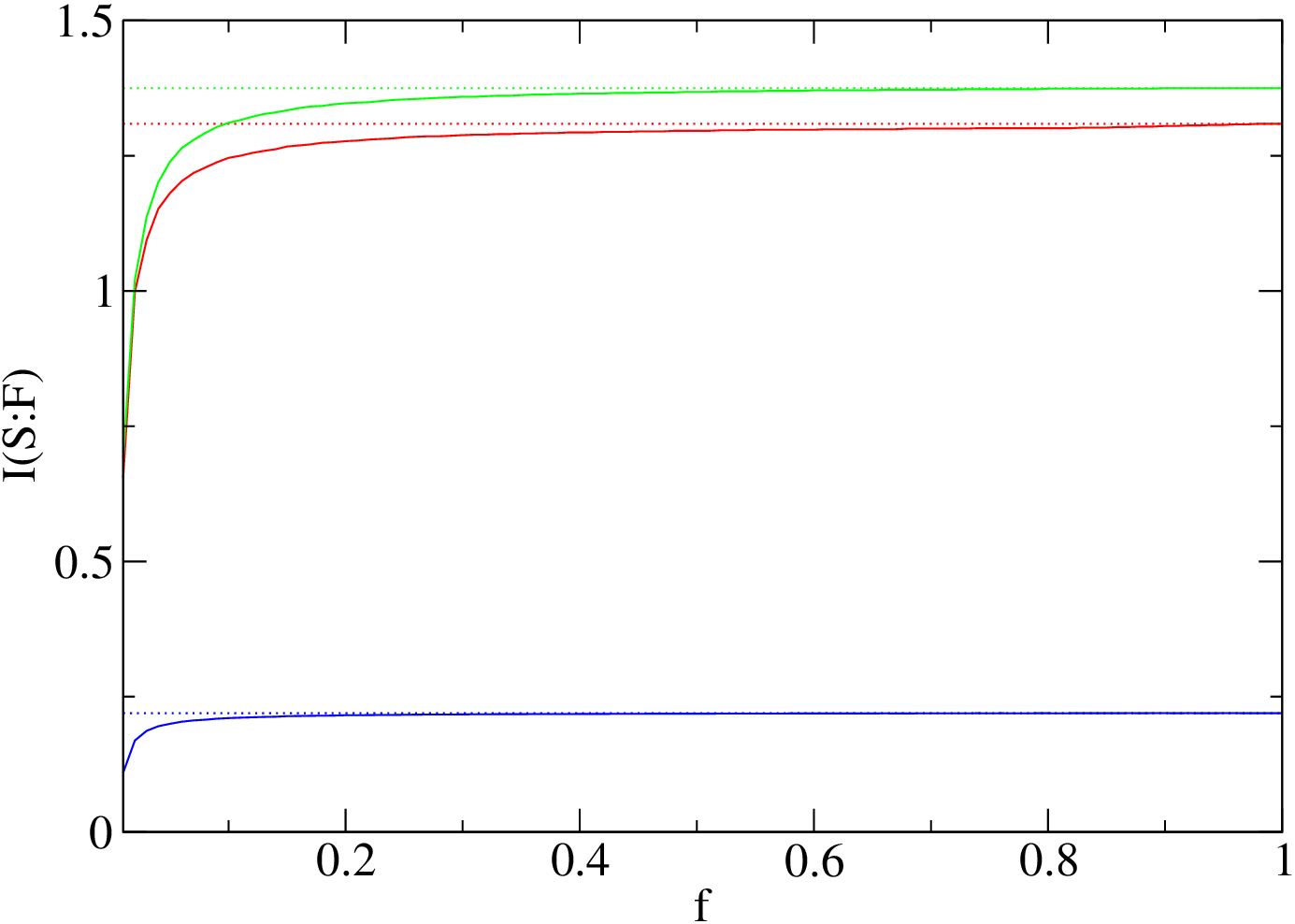}

\caption{\label{n100mutual} (Color online). Same as Fig. \ref{n10mutual}
for the model with $N=100$ .}

\end{figure}

The results for the quantity $I(S:F)$ are shown in Fig. \ref{n100mutual}.
In this case, it becomes impossible to perform an average over all
possible $(\begin{array}{c}
N\\
n_{F}\end{array})$ combinations since, for example, $(\begin{array}{c}
100\\
50\end{array})\approx10^{29}$ . Instead, we have performed an approximated average over a sufficiently
large number of combinations, until we obtain convergence. We observe
that the mutual information converges faster towards its maximum value.
This means that knowing a low fraction of the entire environment will
provide almost complete information about the system. One could also
try to interpret this result in the spirit of \textit{redundant information}
being stored in the environment \cite{Zurek2009,Blume-KohoutPhys.Rev.Lett.101:2404052008,Blume-Kohout2006,Blume-Kohout2005}
(in the sense that many fragments share the same information). It
is interesting to observe that a measurement over virtually any small
random subset of levels configuring our system \textit{E} can be used
to obtain information about \textit{S} . There is also a question
regarding the relevant time scales for the problem. Fig. \ref{n100mutual}
suggests that the time scale necessary for information to be distributed
throughout the environment may be much shorter than the decoherence
time $T_{d}\sim1/\Gamma$ (compare this figure with Fig. \ref{n10and100dat}).
Clearly, more research is necessary to elucidate this issue.

\section{Conclusions}

In this paper, we studied the interaction of a qubit \textit{S} with
an environment \textit{E} consisting on $N$ levels. The initial information
about the qubit is distributed through \textit{E} via the well-known
process of decoherence. We have investigated how much information
about \textit{S} one can obtain by measuring a subset \textit{F} of
$n_{F}$ levels, as a function of the fraction $f=n_{F}/N$ . As a
measure of the amount of information, we used the mutual information
$I(S:F)$ . We found some differences with the partial information
plots that appear when one considers the mutual information of \textit{S}
with a fraction of a given set of quantum systems that define the
environment \cite{Zurek2009,Blume-KohoutPhys.Rev.Lett.101:2404052008,Blume-Kohout2006,Blume-Kohout2005}.
The reason is that \textit{E} is not a bipartite system of F and its
complementary. For example, the plots we obtain do not have the symmetry
properties that appear in the referenced papers although, when properly
averaged over different fragments of the same size, we find that $I(S:F)$
increases monotonically with $f$ . 

Our results show that, when $N$ is increases, even a small fraction
of \textit{E} can give information about S . One would be tempted
to interpret this result in the spirit of \textit{redundant information}
being stored in the environment, in a similar way that Zurek and coworkers
suggest for a multipartite environment, but now applied to a part
of a single quantum system \textit{E} (although possessing a rich
internal structure). Our work can be interpreted as a further step
in the understanding on how information is distributed throughout
the environment. Clearly, more research has to be done in this direction,
but such knowledge can be used, in principle, to a better design of
quantum systems and quantum computers in the presence of decoherence.
Maybe also for a better understanding of how macroscopic observers
perceive quantum systems, as suggested by Zurek et. al.
\begin{acknowledgments}
I would like to acknowledge the comments made by M.C. Bañuls and I.
de Vega during interesting discussions. This work has been supported
by the Spanish Ministerio de Educación y Ciencia through Projects
AYA2007-67626-C03-01 and FPA2008-03373.
\end{acknowledgments}
\bibliographystyle{apsrev}
\bibliography{/home/perez/investig/qcomputing/biblio/books,/home/perez/investig/qcomputing/biblio/qdarwin,/home/perez/investig/qcomputing/biblio/ham,/home/perez/investig/qcomputing/biblio/opensystems}

\end{document}